\documentclass[aps,pre,superscriptaddress,citeautoscript,twocolumn,reprint,longbibliography,floatfix]{revtex4-2}
\usepackage{graphicx,amssymb,amsmath,epsf,bm,cprotect,comment,physics}
\epsfclipon

\newcommand{\unit}[1]{~\mathrm{#1}}

\renewcommand{\eqref}[1]{Eq.\,(\ref{#1})}

\newcommand{\figref}[1]{Fig.\,\ref{#1}}
\newcommand{\figpref}[2]{Fig.\,\ref{#1}(#2)}

\newcommand{\secref}[1]{Sec.\,\ref{#1}}

\newcommand{\etal}{\textit{et al}.\ }

\begin{document}

\title{Various phases of active matter emerging from bacteria and their implications}

\author{Kazumasa A. Takeuchi}
\email{kat@kaztake.org}
\affiliation{Department of Physics,\! The University of Tokyo,\! 7-3-1 Hongo,\! Bunkyo-ku,\! Tokyo 113-0033,\! Japan}%
\affiliation{Universal Biology Institute,\! The University of Tokyo,\! 7-3-1 Hongo,\! Bunkyo-ku,\! Tokyo 113-0033,\! Japan}%
\affiliation{Institute for Physics of Intelligence,\! The University of Tokyo,\! 7-3-1 Hongo,\! Bunkyo-ku,\! Tokyo 113-0033,\! Japan}%

\author{Daiki Nishiguchi}
\affiliation{Department of Physics,\! Institute of Science Tokyo,\! 2-12-1 Ookayama,\! Meguro-ku,\! Tokyo 152-8551,\! Japan}%
\affiliation{Department of Physics,\! The University of Tokyo,\! 7-3-1 Hongo,\! Bunkyo-ku,\! Tokyo 113-0033,\! Japan}%

\date{\today}

\begin{abstract}
In this perspective article, we discuss bacterial populations as a model system of active matter. 
It allows for the exploration and characterization of various phases of active matter and brings rich implications for both physics and biology.
Specifically, we focus on active gas, active liquid, active glass and active liquid crystal states observed in bacterial populations and describe how these differ from their thermal counterparts.
A few future directions are also discussed that will deepen the physical interest in active matter as a new type of material, with its implications for several life phenomena observed in bacterial populations and other biological systems.

\end{abstract}

\maketitle

\section{Introduction}

Active matter has traditionally referred to groups of self-propelled particles, often inspired by living systems such as migrating cells, motor proteins, and animal flocks.
Recently, however, it has become more common to consider that the essence is 
in the fact that particles take free energy or material from their environment and use it for some form of activity, not necessarily self-propulsion \cite{teVrugt.etal-RMP2026}.
Indeed, cells often not only migrate but also grow and divide.
Some enzymes are known to undergo conformational changes during catalysis \cite{Zhang.Hess-ACS2019}.
This generalized definition of active matter has not only allowed a wider range of systems to be explored and characterized within the framework of active matter physics, but also enabled broader questions about life to be discussed in relation to active matter \cite{Hallatschek.etal-NRP2023}.


In this perspective article, we focus on bacterial populations
(also called bacterial active matter \cite{Aranson-RPP2022}) 
as a model experimental system for such general active matter, which has natural biological connections and in which both self-propulsion and growth play major roles.
Physically, bacterial cells often have simple shapes, as represented by the rod morphology of \textit{Escherichia coli} and \textit{Bacillus subtilis}, and deformation is limited in most physiological conditions.
Bacterial motility modes are also extensively studied, despite their diversity across species \cite{Miyata.etal-GC2020,Wadhwa.Berg-NRM2021}.
For instance, \textit{E. coli} and \textit{B. subtilis} are known to exhibit the so-called run-and-tumble motion, i.e., alternating straight runs and random reorientations.
This swimming mode is realized by rotating bacterial flagella, and such a microswimmer can be expressed by an outward force dipole in Stokes hydrodynamics, called pusher, in the lowest-order approximation \cite{Drescher.etal-PNAS2011,Lauga-Book2020,Aranson-RPP2022}.
These simple properties are advantageous when interpreting experimental observations, constructing theoretical models, and so forth. 
Nevertheless, they result in rich collective properties as discussed below. 
Through the combination with sensing, regulations and adaptations, biological functions such as chemotaxis may also emerge.

There are also good biological reasons to focus on bacteria.
Besides taking a major fraction of biomass on the Earth \cite{Bar-On.etal-PNAS2018}, bacteria often live in the form of dense communities, most notably as biofilms \cite{HallStoodley.etal-NRM2004,Flemming.etal-NRM2016,Hallatschek.etal-NRP2023}, i.e., dense aggregates of cells adhering on solid surfaces and covered by extracellular matrix.
Biofilms are a major focus of research across various areas of science, medicine, and industry because they constitute an integral part of ecosystems \cite{HallStoodley.etal-NRM2004} and are implicated in a large fraction of bacterial infections \cite{HallStoodley.etal-NRM2004,Roemling.Balsalobre-JIM2012}, while also finding useful industrial applications in the degradation of organic waste and the production of fine chemicals \cite{Halan.etal-TB2012}. Despite their broad significance, however, satisfactory characterization and control of biofilms remain challenging.
Active matter physics has the potential to provide new approaches and techniques for this area of research, as we briefly discuss in this article.
Another interesting aspect of bacteria lies in quorum sensing \cite{Miller.Bassler-ARM2001,PerezVelazquez.etal-BMB2016}, i.e., a collective response of bacteria to changes in population density, mediated by biochemical signaling and gene regulation.
Besides its biological significance, quorum sensing is known to interplay with flow and geometry \cite{Dalwadi.Pearce-PNAS2021} and may drive collective phenomenon characteristic of active matter, most notably the motility-induced phase separation \cite{Cates.Tailleur-ARCMP2015,Ridgway.etal-PRL2023}.
As such, diverse active matter phases of dense bacterial populations and their properties may naturally have implications on physiological functions of dense bacterial communities.

\begin{figure}[b!]
\centering
\includegraphics[width=\hsize]{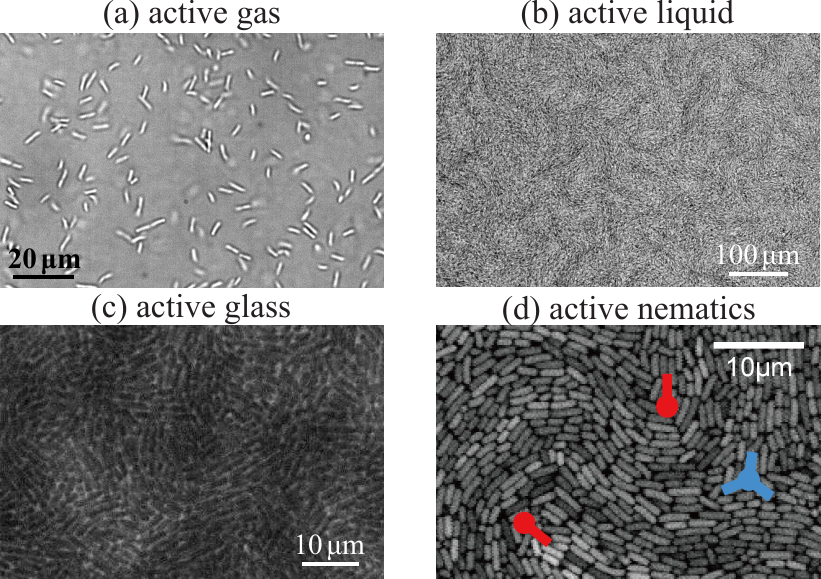}
\caption{
Various phases of active matter observed in bacteria.
(a) Active gas state of swimming \textit{B. subtilis}.
(b) Active liquid state of a concentrated suspension of \textit{B. subtilis} showing turbulent collective motion.
(c) Active glassy state of motile \textit{E. coli} observed in a microfluidic device \cite{Lama.etal-PN2024}.
(d) Active nematic state of nonmotile \textit{E. coli} in a growing colony \cite{Shimaya.Takeuchi-PN2022}.
Images in (c)(d) are adapted from the references cited.
}
\label{fig1}
\end{figure}

With these in mind, here we discuss several active matter phases known to arise in bacterial populations (\figref{fig1}).
The purpose of the article is not to provde a comprehensive review of each phase, but rather to highlight characteristic properties of such active matter phases through comparison with their corresponding thermal phases, and to give some perspectives on the future directions and implications.

\begin{figure}[b!]
\centering
\includegraphics[width=\hsize]{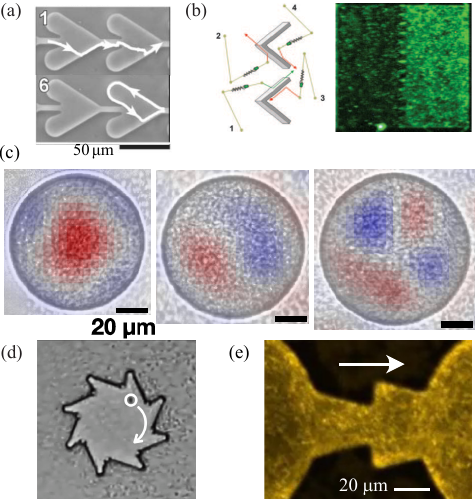}
\caption{
Characteristic transport phenomena in active gas and liquid states of bacteria.
(a) Rectification of a single swimming \textit{E. coli} cell by a ratchet-shaped channel \cite{Hulme.etal-LC2008}.
(b) Spontaneous concentration of \textit{E. coli} suspension induced by an array of funnels \cite{Galajda.etal-JB2007}. The device's inner size was $400\unit{\mu m} \times 400\unit{\mu m}$.
(c) Route to bacterial turbulence as confinement size is varied \cite{Nishiguchi.etal-PNAS2025}. Color indicates the vorticity.
(d) Gear turned by turbulent bacterial suspension \cite{DiLeonardo.etal-PNAS2010,Sokolov.etal-PNAS2010}. The gear diameter is $48\unit{\mu m}$.
(e) Rectification of dense bacterial suspension by a ratchet-shaped channel \cite{Uchida.etal-a2026}.
Images are adapted from the references cited (image in (d) is from \cite{DiLeonardo.etal-PNAS2010}).
}
\label{fig2}
\end{figure}

\section{Active gas and liquid states} \label{sec:fluid}

The gaseous state, i.e., a low-density group of swimming bacteria [\figpref{fig1}{a}], is arguably the simplest phase of bacterial active matter. 
Yet, we may already observe a variety of nontrivial space-time dynamics, in particular when bacteria interact with boundaries of the system.
Attraction to solid surfaces due to collisions and hydrodynamic interactions \cite{Aranson-RPP2022,Berke.etal-PRL2008,Li.Tang-PRL2009,Bianchi.etal-PRX2017,Takaha.Nishiguchi-PRE2023} and surface-induced chirality \cite{DiLuzio.etal-N2005,DiLeonardo.etal-PRL2011} are well-known phenomena.
Moreover, boundaries with asymmetric shapes, such as ratchet-shaped channels and funnels, may induce directional flow of bacterial gas and resulting local concentration of bacteria \cite{Hulme.etal-LC2008,Galajda.etal-JB2007,Reichhardt.Reichhardt-ARCMP2017} [\figpref{fig2}{a,b}].
Such spontaneous rectification and boundary-induced concentration are phenomena that cannot occur in thermal gas, representing distinguished properties of the active gas phase.

This exotic nature of active matter may be even more pronounced in the active liquid state, characterized by strongly correlated motion of active particles.
A common way to realize an active liquid state with swimming bacteria is to culture bacterial suspension, then concentrate it using a centrifuge.
Dense bacterial suspension obtained thereby is known to show a turbulent state spontaneously \cite{Aranson-RPP2022} [\figpref{fig1}{b}].
Indeed, more generally, in such suspensions of microswimmers or active particles, hydrodynamic interactions typically destabilize the uniform flocking state and render it turbulent \cite{Alert.etal-ARCMP2022} (but not always \cite{Nishiguchi.etal-PRE2017,Ghosh.Cheng-PRR2022,Nishiguchi-JPSJ2023}).

Bacterial turbulence, as well as many other examples of active turbulence, consists of vortices with specific sizes \cite{Aranson-RPP2022,Alert.etal-ARCMP2022}, and thus it should be distinguished from fully developed turbulence observed in simple fluids.
In the case of \textit{B. subtilis}, the vortex size is typically about tens of microns, while for \textit{E. coli} it was recently reported that the vortex size scales proportionally with the suspension thickness, reaching even several milimeters \cite{PerezEstay.etal-a2025}.
This difference probably stems from the fact that \textit{B. subtilis} is aerobic and tends to stay near the air-liquid interface, forming an effectively quasi-two-dimentional layer of dense suspension, while oxygen is not as necessary for \textit{E. coli} and they tend to develop a more three-dimensional active suspension.
In any case, the existence of characteristic vortex size implies that flow may be affected and even controlled by geometric designs \cite{Wioland.etal-PRL2013,Beppu.etal-SM2017,Nishiguchi.etal-NC2018,Nishiguchi.etal-PNAS2025}.
Indeed, it was recently shown that the confinement size serves as a control parameter to characterize a route to bacterial turbulence, which starts from regularly circulating flow under strong confinement, passes through a reversing vortex pair, and finally reaches the bulk turbulent state \cite{Nishiguchi.etal-PNAS2025} [\figpref{fig2}{c}].
This route to active turbulence is expected to be generic and is contrasted with standard routes to turbulence of simple fluids \cite{Eckmann-RMP1981}.

From the more practical viewpoint, a striking observation was reported that bacterial active liquid can turn a gear-shaped object [\figpref{fig2}{d}], with fluctuations but on average unidirectionally \cite{Sokolov.etal-PNAS2010,DiLeonardo.etal-PNAS2010} .
For thermal liquid, such a phenomenon would be clearly forbidden by the second law of thermodynamics, but it does not violate anything in the case of active liquid, at the price of energy dissipated everywhere in the liquid.
Rectification was also shown to occur in a ratchet-shaped channel \cite{Uchida.etal-a2026} [\figpref{fig2}{e}].
The realization and exploration of such phenomena that are not possible for thermal systems underscores the significant potential of active matter as an exotic material.

Theoretically, there are several approaches to describe such active gas and liquid states.
If suspension is viewed as a group of self-propelled particles, microswimmer models \cite{Lauga-Book2020}, based on Stokes hydrodynamics appropriate for such low-Reynolds-number systems, can account for many of the cell-cell and cell-surface interactions observed in bacteria \cite{Aranson-RPP2022}.
If hydrodynamic interactions do not play a major role, so-called dry models may suffice.
A variety of models of self-propelled particles have been used in the literature, featuring different aspects: roughly, the Vicsek model for alignment interactions \cite{Chate-ARCMP2019}, the active Brownian particle model for positional interactions \cite{Bechinger.etal-RMP2016}, the run-and-tumble model for tumbling behavior of swimming bacteria \cite{Bechinger.etal-RMP2016}, and self-propelled rod models for steric interactions of rod-shaped bacteria \cite{Baer.etal-ARCMP2019}, to name but a few.
Complementary to these agent-based models are continuum descriptions.
Continuum equations have been formulated for different symmetry classes \cite{Chate-ARCMP2019,Baer.etal-ARCMP2019}: the polar active fluid, the apolar (nematic) active fluid, and the case for self-propelled rods, with polar self-propulsion and apolar interactions.
A subtlety lies in the fact that rod-shaped bacteria generally have mixed symmetries: they self-propel directionally, but some bacteria are known to show reversal motion \cite{Wadhwa.Berg-NRM2021}.
The steric interaction between cell bodies is essentially apolar, but polarity may emerge when flagella are taken into account \cite{Aranson-RPP2022}.
Hydrodynamic interaction may also contribute to polarity, although the flow generated by a swimming bacterial cell, which typically behaves as a pusher, is predominantly apolar to leading order in the far field \cite{Drescher.etal-PNAS2011,Lauga-Book2020,Aranson-RPP2022}.
It is therefore crucial to choose a description appropriate for the phenomenon of interest.
For example, bacterial turbulence may be described phenomenologically by a polar active fluid equation, supplemented by the Swift-Hohenberg-type instability for vortex formation \cite{Aranson-RPP2022,Alert.etal-ARCMP2022}.
Bacteria also form clusters of collectively propelling cells \cite{Zhang.etal-PNAS2010,Peruani.etal-PRL2012} and a bidirectional flocking state with long-range orientational order \cite{Nishiguchi.etal-PRE2017,Ghosh.Cheng-PRR2022,Nishiguchi-JPSJ2023}, for which self-propelled rod models may give more appropriate descriptions \cite{Baer.etal-ARCMP2019}.

\begin{figure}[b!]
\centering
\includegraphics[width=\hsize]{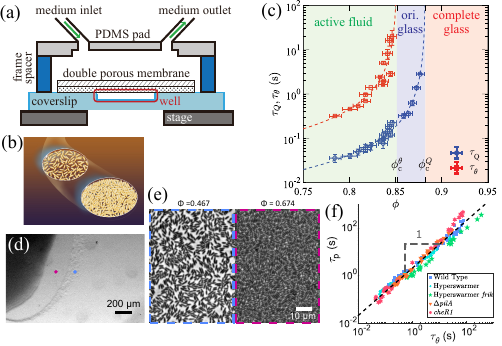}
\caption{
Transitions to glassy states of bacterial populations.
(a) Sketch of the membrane-type microfluidic device used by Lama \etal \cite{Lama.etal-PN2024}. Adapted from \cite{Shimaya.etal-CP2021}.
(b) Sketch of the bacterial glass transition occurring in the course of time in a microfluidic well. Credit: Tomo Narashima.
(c) Two-step transition of motile \textit{E. coli} in a microfluidic well reported by Lama \etal \cite{Lama.etal-PN2024}. Adapted from \cite{Lama.etal-PN2024}.
(d) Tip of a colony of \textit{P. aeruginosa} made by deposition of a water droplet, analyzed by Maliet \etal \cite{Maliet.etal-a2025}.
(e) Close-ups of cell populations at the locations indicated in (d).
(f) Absence of the two-step transition reported by Maliet \etal \cite{Maliet.etal-a2025}.
Panels (d)-(f) are adapted from \cite{Maliet.etal-a2025}.
}
\label{fig3}
\end{figure}

\section{Transitions to glassy states}

As stated earlier, self-propulsion may not be the only active aspect of constituent particles of active matter.
Growth and division, and resulting proliferation, may also play a significant role.
The combination of these has led to the observation of glass transitions in bacterial active matter.

To date, besides an earlier indication \cite{Takatori.Mandadapu-2020}, there are two quantitative experiments of bacterial glass transitions \cite{Lama.etal-PN2024,Maliet.etal-a2025}.
Lama \etal \cite{Lama.etal-PN2024} used a membrane-type microfluidic device \cite{Shimaya.etal-CP2021} that allows for uniform culturing of bacteria in a closed two-dimensional well [\figpref{fig3}{a}].
Swimming bacteria, specifically \textit{E. coli}, proliferated in the well, and eventually became unable to move [\figpref{fig3}{b}].
Through observations of the phase-contrast intensity field and the orientation field, the authors evaluated relaxation times of the intensity change and the orientation change.
The two relaxation times turned out to be significantly different, with the orientational one [red symbols in \figpref{fig3}{c}] being one order of magnitude (or more) larger than the other (blue symbols), and this led the authors to interpret that the intensity change is governed by a degree of freedom other than the orientation, which is presumably the translation. 
Moreover, the orientational relaxation time was found to diverge at a packing fraction lower than the translational one does.
In other words, the glass transition takes place by two steps: at the first transition point, only the orientational degrees of freedom are frozen, while the translational ones remain active until the second transition point [\figpref{fig3}{c}].
The three regions separated thereby correspond to the active fluid, orientation glass, and complete glass states.
Interestingly, the orientation glass state appears analogous to the so-called liquid glass state predicted theoretically for anisotropic particles \cite{Letz.etal-PRE2000}.
Previously, the orientation glass state was also reported for passive colloidal rods \cite{Zheng.etal-PRL2011,Zheng.etal-NC2014,Mishra.etal-PRL2013}, so that Lama \textit{et al.}'s observation can be regarded as its active counterpart.

This conclusion was however contrasted by Maliet \textit{et al.}'s experiment on \textit{Pseudomonas aeruginosa} swarming at colony edges \cite{Maliet.etal-a2025}.
More precisely, the authors used edge regions of colonies on agar, which were monolayers of nearly immotile bacteria.
By dropping a water droplet, cells were dispersed as water was absorbed to gel and cells also migrated from the colony bulk.
This made regions of swarming bacteria at different packing fractions [\figpref{fig3}{d,e}], which the authors analyzed.
The authors used a deep-learning method ``DiSTNet2D'' \cite{Ollion.etal-PL2024} to segment single cells and track them, and analyzed both the translation and orientation of single cells.
Then, remarkably, their relaxation times turned out to be proportional to each other for all packing fractions [\figpref{fig3}{f}].
It follows that the two degrees of freedom diverge at the same packing fraction, indicating the absence of the two-step transition.
The authors also genetically engineered the strain and reached the same conclusion for all the mutants [different colors in \figpref{fig3}{f}].

How can we interpret these apparently conflicting conclusions of the two experiments?
On the one hand, there is no conflict, as the two experiments used different bacterial species, which are known to have completely different motility modes \cite{Wadhwa.Berg-NRM2021}, as well as different experimental setups and different analysis methods.
On the other hand, it is always insightful to consider what made the different results.
Among the differences mentioned above, it may be worth noting that Lama \etal used field variables and Maliet \etal used single cell quantities; in other words, they used Eulerian and Lagrangian variables, respectively.
Whereas this difference is likely to be irrelevant in the normal glassy state, in the orientation glass, cells may still change positions $\bm{r}_i(t)$ while their orientations $\theta_i(t)$ passively follow the frozen orientation field $\theta(\bm{r})$, $\theta_i(t) = \theta(\bm{r}_i(t))$, as suggested from observed trajectories \cite{Lama.etal-PN2024}.
This suggests that, even if the relaxation time of the orientation field diverges, that of single cell orientations can remain finite and be essentially determined by that of the translation.
This hypothesis, if confirmed by careful inspection, may give an important hint to interpret the two experimental results.

More broadly, such active glass has several aspects that are worth discussing and investigating.
Physically, it is important to understand how the activity changes characteristic properties of glass known from thermal systems.
This point has indeed been widely studied in the literature \cite{Janssen-JPCM2019}, mainly through numerical and theoretical approaches, and it turned out that the effect of the activity is nontrivial, sometimes contributing to fluidization and other times to vitrification.
Otherwise, experimentally, Lama \etal reported the appearance of an anomalous peak in the dynamic susceptibility at low densities, which was interpreted to result from collective motion of bacteria \cite{Lama.etal-PN2024}.
This can be generic in active rods but is yet to be tested.

Biologically, it is noteworthy that physics of glass seems to be relevant in diverse contexts.
For example, the interior of a living cell is a crowded environment filled with various biomolecules.
While attempting to achieve the same concentration in a cell extract would cause it to vitrify, in a living cell, its metabolic activity was found to maintain the fluidity of the cytoplasm \cite{Janssen-JPCM2019,Nishizawa.etal-SR2017}.
Epithelial cell tissues constitute another example where the relationship to glass has been discussed \cite{Janssen-JPCM2019}.
After all, the characteristic of glass that even slight changes in density can drastically alter the timescale of dynamics may be utilized by life in various ways. 
For example, there was a report that tardigrades utilize vitrification of disordered proteins to protect themselves in arid environments \cite{Boothby.etal-MC2017}. 
This breadth of examples implies that developing the physics of active glass may contribute to understanding a variety of life phenomena.

\section{Active liquid crystal}

The active counterpart of liquid crystalline states, in particular active nematics, has witnessed remarkable development in the last decade or so \cite{Doostmohammadi.etal-NC2018,Doostmohammadi.Ladoux-TCB2022}. 
Examples range from cytoskeletal filaments to cell sheets. 
Even the relevance to morphogenesis upon \textit{Hydra} regeneration has been discussed \cite{Maroudas-Sacks.etal-NP2021,Ravichandran.etal-SA2025}.

Rod-shaped bacteria, such as \textit{E. coli}, \textit{Myxococcus xanthus}, and \textit{Vibrio cholerae}, also develop nematic ordering in dense aggregates [\figpref{fig1}{d}], and as such the relevance to colony growth \cite{Doostmohammadi.etal-PRL2016,Copenhagen.etal-NP2021,Shimaya.Takeuchi-PN2022} and biofilm architecture \cite{Drescher.etal-PNAS2016,Hartmann.etal-NP2018,Yaman.etal-NC2019,Zhang.etal-PNAS2021,Nijjer.etal-NP2023,Prasad.etal-S2023} has been extensively studied.
Physically, activity of cells, whether motility or growth, can be expressed as force dipoles along rod particles, and this gives rise to active stress proportional to the nematic tensor order parameter \cite{Doostmohammadi.etal-NC2018}, a term forbidden for thermal systems.
This renders nematic topological defects interesting features, through characteristic stress patterns around defects, such as self-propulsion of $+1/2$ defects \cite{Giomi.etal-PRL2013,Doostmohammadi.etal-NC2018}. 

\begin{figure}[t!]
\centering
\includegraphics[width=\hsize]{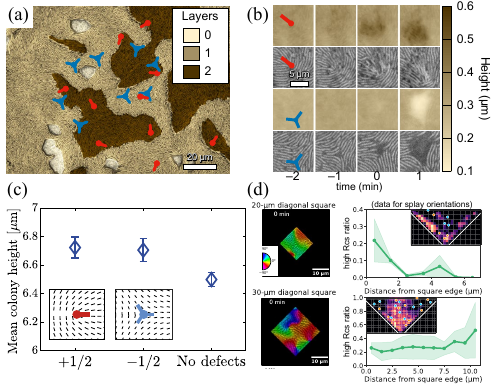}
\caption{
Roles of nematic order in bacterial populations.
(a)(b) Topological defects in populations of \textit{M. xanthus} (a) and time changes around $+1/2$ and $-1/2$ defects [(b) top and bottom, respectively] \cite{Copenhagen.etal-NP2021}.
(c) Mean local heights of \textit{E. coli} colonies at the location of defects in the bottom layer [an example shown in \figpref{fig1}{d}] compared with that in the region without defects \cite{Shimaya.Takeuchi-PN2022}.
(d) Topological defects promote the production of colanic acid in \textit{E. coli} populations \cite{Yokoyama.Takeuchi-b2026}. Cell orientations (left), the locations of $+1/2$ and $-1/2$ defects (right insets, orange and blue disks, respectively) and the high Rcs activation density indicating the spatial distribution of the colanic acid production (right insets pseudocolor) are compared between two square-shaped confiner devices (diagonal lengths $20\unit{\mu m}$ and $30\unit{\mu m}$ for top and bottom panels, respectively). The data in the right insets were assembled from multiple samples. The right main panels show the mean fraction of high Rcs pixels against distance from the square edge. In the smaller device, when splay orientation patterns are formed around the corners, $+1/2$ defects are found predominantly there (top right inset, orange disks) and, as a result, the colanic acid production tends to be higher near the edges (top right, green curve). 
In the larger device, defect positions are more distributed in the bulk (bottom right inset, disks) and the colanic acid production becomes higher in the bulk (bottom right, green curve). Figures are adapted from the references cited.
}
\label{fig4}
\end{figure}

Combined with anisotropic frictions, defects in active nematics are also known to attract or repel cells  \cite{Kawaguchi.etal-N2017,Copenhagen.etal-NP2021,Shimaya.Takeuchi-PN2022}.
In the case of bacteria, this was clearly observed in populations of \textit{M. xanthus}, where comet-shaped $+1/2$ defects attracted cells and trefoil-shaped $-1/2$ defects repelled cells, triggering the formation of an upper layer of cell sheets for the former and holes for the latter \cite{Copenhagen.etal-NP2021} [\figpref{fig4}{a,b}].
This attraction and repulsion of cells for $+1/2$ and $-1/2$ defects is consistent with an earlier observation for neural progenitor cells \cite{Kawaguchi.etal-N2017} and was successfully explained theoretically on the basis of two-dimensional active nematics \cite{Kawaguchi.etal-N2017,Copenhagen.etal-NP2021}.
However, while bacteria grow two-dimensionally at the early stage of colony development, they start growing three-dimensionally when pressure builds up \cite{Beroz.etal-NP2018,You.etal-PRL2019}.
For biofilm-forming species, this three-dimensional growth is a precursor to the development of biofilms \cite{Drescher.etal-PNAS2016,Beroz.etal-NP2018,Yaman.etal-NC2019}.
Nematic patterns also provide key information for elucidating well-developed three-dimensional morphologies of biofilms \cite{Drescher.etal-PNAS2016,Zhang.etal-PNAS2021,Nijjer.etal-NP2023,Prasad.etal-S2023}.
In three dimensions, topological defects take the form of lines and, even if cross sections are taken, $+1/2$ and $-1/2$ defects are topologically equivalent and intermediate structures are permitted.
The relevance of such line defects to the biofilm formation started to be discussed \cite{Zhang.etal-PNAS2021,Nijjer.etal-NP2023}.
It was also reported that, even in weakly three-dimensional colonies, cell tilting can generate additional forces that act inward to $-1/2$ defects \cite{Shimaya.Takeuchi-PN2022}.
As a result, cells are attracted to both $+1/2$ and $-1/2$ defects and raise the local height of the colonies slightly but significantly [\figpref{fig4}{c}].

The influence of topological defects may not be limited to colony morphologies.
Biologically, the expression of certain genes is known to be induced by mechanical stimuli through a process known as mechanotransduction.
This can be triggered by active stresses associated with topological defects, as reported for mammalian epithelial cell tissues \cite{Saw.etal-N2017}.
Recently, an analogous phenomenon was suggested in bacteria, specifically involving the production of colanic acid \cite{Yokoyama.Takeuchi-b2026} [\figpref{fig4}{d}], a molecular component of the biofilm extracellular matrix, known to protect cells from phages and also sustain the three-dimensional architecture of biofilms \cite{Limoli.etal-MS2015}.

The relevance of active liquid crystals to life paves the way for controlling various life phenomena using techniques developed in liquid crystal science.
Regarding the colanic acid, it was shown that the distribution of defects can be changed by the design of microfluidic devices, which then changes the distribution of produced colanic acid \cite{Yokoyama.Takeuchi-b2026} [\figpref{fig4}{d}].
Since the extracellular matrix is an indispensable component of biofilms with characteristic mechanical properties \cite{Krishnan.etal-BR2025,Yan.etal-AM2018}, serving numerous functions including mechanical support of biofilm architecture, signal transduction, cell protection, extracellular metabolic activities \cite{HallStoodley.etal-NRM2004,Flemming.etal-NRM2016}, the possibility of controlling the extracellular matrix through active-nematic mechanisms may offer a promising avenue for biofilm control.
Beyond bacteria, for \textit{Hydra}, compression of tissues may cause different topological arrangements and result in different body shapes \cite{Ravichandran.etal-SA2025}.
More generally, various alignment techniques have been developed in liquid crystal science.
By using these, one can design the orientation pattern of cell populations and thereby place topological defects at desired positions \cite{Doostmohammadi.Ladoux-TCB2022}.
It is of significant importance to use these techniques to control life phenomena linked to cellular or cytoskeletal orientations.
An important challenge is extending the theoretical understanding and experimental techniques to the three-dimensional case, which would further enrich this line of research.

Finally, it is worthwhile to remark, following \secref{sec:fluid}, that polar properties of bacteria may sometimes be important.
It was recently reported that local polar order in self-propulsion plays a significant role in the defect-mediated layer formation of \textit{M. xanthus} [\figpref{fig4}{a)(b}] \cite{Han.etal-NC2025}.
Even in nonmotile bacteria, polar effects may emerge through asymmetric adhesion of cells \cite{Duvernoy.etal-NC2018}.
In such cases, polar active rod models may complement active nematic theories for better descriptions of the phenomena.


\section{Perspectives}

In this perspective article, we have featured bacterial populations as a model system to explore and characterize different phases of active matter.
We have discussed how each phase exhibits characteristics distinct from those of thermal counterparts, and how their potential as new materials and their relationship to life phenomena are beginning to emerge.

When considering how active matter can ultimately contribute to physics and life sciences, as physicists, the present authors find an analogy to what Philip W. Anderson discussed in his essay entitled ``More is Different'' \cite{Anderson-S1972}.
As the present authors interpret, he pointed out in the essay that both the understanding of constituent particles and that of emergent phenomena -- emerging as a result of interactions of constituent particles -- are important.
This view has been supported by the remarkable, continuing advancements in both high-energy and condensed matter physics.
Now, in life sciences, the glorious development of molecular biology has certainly constructed the solid understanding and surprising beauty of life phenomena, on the basis of constituent genes and biomolecules.
Then, what about the ``More is Different'' aspect?
While cooperative phenomena of genes and molecules have been of course studied in molecular biology and other disciplines too, it is not unreasonable to consider that active matter physics may provide one approach to deal with emergent phenomena in living systems.
Similarly to condensed matter physics, which has both fundamental and applied sides with various interesting directions, active matter physics also certainly has both fundamental and applied interests, as a novel materials science and as a ``More is Different'' approach for life, among others.

\begin{acknowledgments}
The authors' studies discussed in this article were the result of many collaborative efforts and discussions with various colleagues.
We thank in particular Igor S. Aranson, Hugues Chat\'e, Masaki Sano, Andrey Sokolov, and Yuichi Wakamoto, as well as past and present members of Takeuchi Group, namely Yujiro Furuta, Hisay Lama, Takuro Shimaya, Sora Shiratani, Yoshihito Uchida, Masahiro Yamamoto, and Fumiaki Yokoyama.
This work is supported by KAKENHI from Japan Society for the Promotion of Science (JSPS) (Nos. JP24K00593, JP19H05800, JP16H04033, JP26H00388 to K.A.T. and Nos. JP25K22005, JP23H01141, JP20K14426, JP19K23422, JP26H00387 to D.N.), Japan Science and Technology Agency (JST) FOREST (Nos. JPMJFR2364 to K.A.T. and JPMJFR256R to D.N.) and PRESTO (No. JPMJPR21O8 to D.N.), and the JSPS Core-to-Core Program ``Advanced core-to-core network for the physics of self-organizing active matter (No. JPJSCCA20230002).''
\end{acknowledgments}

\bibliography{ref}

\end{document}